\documentclass[prd,nofootinbib,preprintnumbers,showpacs,floats,aps,twocolumn]{revtex4}
\usepackage{graphicx,xspace,units}
\usepackage{float}
\usepackage{amsmath}
\usepackage{amssymb}
\usepackage[normalem]{ulem}
\usepackage{wrapfig}
\usepackage{epsfig}
\usepackage{bm}
\usepackage[usenames]{color}
\usepackage{siunitx}
\usepackage{ulem}

\newcommand{\eq}[1]{eq.~(\ref{#1})}
\newcommand{\Eq}[1]{Eq.~(\ref{#1})}
\newcommand\brak[1]{\ensuremath{\bigl| #1\bigr\rangle}}
\newcommand\krab[1]{\ensuremath{\bigl\langle #1\bigr|}}
\newcommand\brakket[2]{\ensuremath{\,\bigl\langle #1 \big| #2\bigr\rangle\,}}
\begin{document}

\preprint{MIT-CTP/5253}
\title{Ambiguities in the definition of local spatial densities in light hadrons}
 \today
\author{R. L. Jaffe}
\email{jaffe@mit.edu}
\affiliation{Department of Physics,  Center for Theoretical Physics, and
Laboratory for Nuclear Science,
Massachusetts Institute of Technology,
Cambridge, MA 02139,
USA}

\begin{abstract}   
 The relationship between the matrix element of a local operator and the Fourier transform of the associated form factor fails for systems such as the nucleon where its intrinsic size is of order its Compton wavelength.  Although one can conceive of an intrinsic charge density distribution in the proton, there does not seem to be an unambiguous way to define, compute, or measure it precisely.
\end{abstract}
\maketitle


\section{Introduction} 

This note explores and quantifies an impediment to defining local densities in ordinary 3-space in systems whose size is of the same order as their Compton wavelength.  A classic example is the electric charge density distribution in the nucleon $\rho_N(r)$. It is widely, but erroneously, believed  that $\rho_{N}(r)$ is given by the Fourier transform of the nucleon's electric form factor.   
This identification was proposed long ago\cite{hofstadter,ernst,sachs} as an extension of the well-known relation that holds for non-relativistic systems like atoms.  The ``derivation'' for the nucleon case is usually credited to Sachs\cite{sachs}.  More recently, analogous relations have been proposed between Fourier transforms of the form factors of components of the energy momentum tensor and other local densities such as the pressure and shear force within nucleons\cite{Polyakov:2002wz,Polyakov:2002yz,Polyakov:2018zvc}.

This identification is certainly valid for non-relativistic systems such as atoms and also, with relatively small ambiguities, for nuclei.  It fails badly, however, for a system  like the nucleon whose Compton wavelength ($0.21$~fm) is comparable to its size ($0.85$~fm).  Furthermore the problem is more general than the charge density distribution.  Indeed, it does not seem to be possible to unambiguously define, compute, or measure the spatial dependence of the nucleon matrix element of any local operator independent of the specific form of the wave packet state in which the nucleon was prepared.  The same conclusion applies to other light hadrons or, in fact, any system for which the Compton wavelength and intrinsic size are comparable. 

This problem has been known, but not widely known, for many years\cite{Burkardt:2000za}\footnote{I first became aware of this problem through discussions with M. Burkardt, see Ref.~\cite{Burkardt:2000za}.}.  Miller, in particular, has stressed the problem of identifying the nucleon charge distribution $\rho_{N}(r)$ with the Fourier transform of the electric form factor\cite{Millerold}.   Recent lattice calculations of the quark and gluon pressure and other distributions in hadrons reminded me of the problem and stimulated this note.   In a recent paper\cite{Miller:2018ybm} Miller has again called attention to the problem and in particular to the failure of the widely accepted connection between the nucleon's charge radius and the derivative of its electric form factor at zero momentum transfer, $r_p^2=-6 G_E'(0)$.\footnote{For further references see \cite{Miller:2018ybm}.}  Miller demonstrates that this relation is not valid when sufficient care is taken to localize the nucleon and keep track of relativistic effects.  In a sense, this note can be considered a further exploration of the arguments put forward in Refs.~\cite{Burkardt:2000za}
 and \cite{Miller:2018ybm}, emphasizing that the problem can be appreciated with only an elementary knowledge of relativistic quantum mechanics and illustrating it in a simple model.\footnote{Incidentally, the problem does not apply to the matrix elements of \emph{bilocal} operators like those that define (ordinary, generalized, and transverse momentum dependent) parton distribution functions.  As first recognized by Soper\cite{Soper:1976jc} and emphasized by Burkardt\cite{Burkardt:2000za} (see also Refs.~\cite{Millerold,Miller:2018ybm}) these correlation functions can be manipulated to define local distribution functions in the plane transverse to the direction defined by the infinite momentum frame of the parton model\cite{lorce}.}  

The basic problem is that to measure the matrix element of a local operator like the charge density $\hat\rho({\bf r})$ in any quantum system, the coordinate ${\bf r}$ that appears in the operator has to be defined relative to the location of the system.  Thus the system must be localized somewhere.  It is necessary to construct a localized wave packet whose center defines the coordinate origin  with respect to which ${\bf r}$ is defined.  The more tightly one tries to localize the system in order to give a precise meaning to ${\bf r}$, the higher the momentum components one introduces into its wave function and the larger the relativistic effects that make the matrix element of the operator dependent on the form of the wave packet.  At the other extreme, if one chooses a wave packet that is large compared to the intrinsic size of the system, then the calculated or measured charge distribution is dominated by the width of the wave packet, not by the intrinsic charge distribution of the system.

A detailed analysis in a simple model (see below) shows that troubles arise if one attempts to localize a system within its Compton wavelength.  For an atom, the Compton wavelength is of order $ 0.2/A$~fm while the charge is distributed over several Angstroms, so there is no problem.  For a nucleus, the Compton wavelength (again $0.2/A$~fm) is relatively small compared to its size ($R_{A}\sim 1.3A^{1/3}$~fm), and the ambiguities in the definition of the charge density and other local distributions are not large, except perhaps for the deuteron.  For the nucleon the ambiguities are significant and for the pion they are overwhelming.  The condition for the validity of the traditional Fourier transform connection between the form factor and charge density distribution of an object is
\begin{equation}
\Delta \gg R\gg 1/m\,,
\label{eq:diffresult}
\end{equation}
where $1/m$ is the Compton wavelength of the object, $R$ is a measure of the localization provided by the wave packet, and $\Delta$ is a measure of the intrinsic size of the system, for example $-6 G_{E}'(0)$.  Only when $\Delta \ggg 1/m$ can the $R$ dependence be dismissed.

The textbook analysis of the charge distribution of atoms and nuclei finesses this problem by transforming the Schr\"odinger wave function to center of mass and relative coordinates, thereby defining ${\bf r}$ relative to the center of mass.  A Schr\"odinger wave function description of a relativistic bound state like the nucleon does not exist and the transformation to relative and center of mass coordinates  is not possible.  

The matrix elements of local operators in the nucleon have been calculated for many years.  When these calculations are performed directly in coordinate space, as for example in the MIT bag model, their physical significance is unclear for the reasons just explained.  On the other hand, the calculation of the form factors of local operators in momentum space is unambiguous --- they are in principle measurable in scattering experiments --- as, for example in recent lattice calculations of the pressure and shear force in the nucleon\cite{Shanahan:2018nnv}. The Fourier transform  of such a form factor cannot, however, be interpreted as the coordinate space matrix element of the operator of interest.

Note that this ambiguity has no consequences for the currently interesting discrepancy between different measurements of the nucleon ``charge radius'' since, as Miller and others have pointed out, what is actually measured and disagrees among experiments is the derivative of the nucleon's charge form factor at zero momentum transfer, $G_{E}'(0)$. It is the association of $G_{E}'(0)$ with the nucleon's charge radius that is unwarranted and is further explored in this note.  

\section{Defining the charge density distribution for a quantum particle}

\subsection{Basic definitions}
Spin plays no special role in this analysis, so nothing is lost by considering a spin-0 system.  I have in mind a spinless ``nucleon'', but the analysis applies to any localizable quantum system.  It also does not matter what are the constituents of this system.  Consider the electric charge density operator  $\hat\rho({\bf r},0)$ at $t=0$ in the Heisenberg picture. Suppose the system is an eigenstate of $\hat Q=\int d^3 r \hat\rho({\bf r})$, with eigenvalue $Q$,
\begin{equation}
\hat Q\brak{p}=Q\brak{p}\,.
\label{eq:Q}
\end{equation}
For definiteness, choose $Q=1$.
Here $\brak{p}$ is a covariantly normalized momentum eigenstate,
\begin{equation}
\brakket{p'}{p}=2E(2\pi)^3\delta^3({\bf p}'-{\bf p})\,,
\label{eq:cv}
\end{equation}
where $p=(E,{\bf p})$ and $E=\sqrt{m^{2}+{\bf p}^{2}}$. 

The ${\bf r}$ dependence of the matrix element of $\hat\rho({\bf r},0)$ between momentum eigenstates is determined by the translation invariance of momentum eigenstates,
\begin{equation}
\krab{p'}\hat\rho({\bf r},0)\brak{p}=e^{i({\bf p}'-{\bf p})\cdot {\bf r}}
\krab{p'}\hat\rho(0)\brak{p}\,.
\label{eq:mom-eigstate}
 \end{equation}
Having chosen a spin-0 system, the matrix element in the previous equation is determined by a single  charge form factor, $F(q^{2})$, 
 \begin{align}
\krab{p'}\hat\rho(0)\brak{p}&=(E+E')F(q^2)\,\,{\rm where}\\
q^2=(p'-p)^2&=(E'-E)^2-({\bf p}'-{\bf p})^2\,.
\label{eq:ff}
\end{align} 
Combining these equations we have
\begin{equation}
\krab{p'}\hat\rho({\bf r},0)\brak{p}=e^{i({\bf p}'-{\bf p})\cdot {\bf r}}
 (E+E')F(q^2)\,.
\label{eq:momff}
 \end{equation}
 Nothing further can be done without constructing a wave packet state localized at some position with respect to which the coordinate ${\bf r}$ is defined.
 
 \subsection{The charge density distribution in a  wave packet state}
 
Let us superpose energy-momentum eigenstates to define a localized, Heisenberg picture state for the particle of interest,
\begin{equation}
\brak{\Psi,{\bf x}}= \int \frac{d^3 p}{\sqrt{2E(2\pi)^3}} \phi({\bf p})
e^{-i{\bf p}\cdot {\bf x}}\brak{p},
\label{eq:wavepack}
\end{equation}
which is normalized to one by requiring
\begin{equation}
\int d^3p |\phi({\bf p})|^2 = 1\,.
\label{norm}
\end{equation} 
To localize the particle at the origin, set ${\bf x}=0$ and define $\brak{\Psi,0}\equiv \brak{\Psi}$.  

Although any localized wave packet would do, I choose a spherically symmetric gaussian packet to simplify subsequent calculations, 
\begin{equation}
\phi({\bf p})\equiv \phi( p) =\left(\frac{2R^2}{3\pi}\right)^{3/4}e^{-   {\bf p}^2 R^2/3}\,,
\label{eq:gauss}
\end{equation}
where I have defined the length scale $R$ equal to the  RMS radius of the wave packet.

Then the object of interest is the charge density distribution in the localized  state, 
\begin{equation}
\rho(r)\equiv \krab{\Psi}\rho({\bf r},0)\brak{\Psi}\,.
\label{charge-dist}
\end{equation}
It is obtained by substituting from  \eq{eq:gauss} for $\phi(p)$ and from \eq{eq:momff} for the matrix element between momentum eigenstates,
\begin{equation}
\rho(r) 
=\int \frac{d^3p\, d^3p'}{(2\pi)^3\sqrt{4EE'}}(E+E') 
 F(q^2)\phi(p')\phi(p)
e^{i{\bf q}\cdot {\bf r}}\,.
\label{eq:basic}
\end{equation}
Here ${\bf q}={\bf p}'-{\bf p}$ and $q^{2}=(E'-E)^{2}-{\bf q}^{2}$, and $q^{2}\le 0$.

To further simplify computations I assume that the form factor, $F(q^{2})$ is also a  gaussian parameterized by a length scale $\Delta$,
\begin{equation}
F(q^{2})=e^{\frac{1}{6}q^{2}\Delta^{2}}\,.
\label{eq.ff-gauss}
\end{equation}
The normalization, $F(0)=1$, is chosen so that $\krab{\Psi}\hat Q\brak{\Psi}=1$.  The  size  of the system is parameterized by the naive mean-square charge radius, $r_{\rm naive}^{2}= 6(dF/dq^{2}|_{q^{2}=0}) =\Delta^{2}$, which for the nucleon is approximately (0.85~fm)$^{2}$ .

To proceed, change to relative and total center-of-momentum variables,
\begin{align}
{\bf p} & = {\bf P} + {\bf q}/2\nonumber\\
{\bf p}' & = {\bf P} -{\bf q}/2\, ,
\label{eq:rel}
\end{align}
giving
\begin{align}
\rho(r)&= \left(\frac{2R^{2}}{3\pi}\right)^{3/2}\int \frac{d^3P\, d^3q}{(2\pi)^3\sqrt{4EE'}}(E+E') \nonumber\\ &\times\exp{\left(\tfrac{1}{6}q^{2}\Delta^{2}  -\tfrac{2}{3}{\bf P}^{2}R^{2} 
 -\tfrac{1}{6}{\bf q}^{2}R^{2}+i{\bf q}\cdot{\bf r}\right)}\,,
 \label{eq:cmrel}
\end{align}
where I have substituted from \eq{eq:gauss} for $\phi(p)$.
 
\subsection{Evaluating $\rho(r)$ when $\Delta \ggg 1/m$}
The integral of \eq{eq:cmrel} must be evaluated numerically for systems like the nucleon for which \eq{eq:diffresult} is not satisfied.  On the other hand, for  systems like atoms and nuclei, where \eq{eq:diffresult} holds, it is useful to expand the terms in $\rho(r)$ in inverse powers of $m$.  In particular, we expand the kinematic factor $(E'+E)/\sqrt{4EE'}$ and the form factor $F(q^{2})$ keeping the first significant term in each.   Expansion of the kinematic factor $(E'+E)/\sqrt{4EE'}$ yields,
\begin{equation}
\frac{E'+E}{\sqrt{4EE'}} = 1+\frac{1}{2m^{4}}({\bf P}\cdot {\bf q})^{2}
+{\cal O}(1/m^{6})\,.
\label{eq:kin}
\end{equation}
The nucleon mass enters the form factor through the energy difference,
\begin{align}
q^{2}&=(E'-E)^{2}-{\bf q}^{2} \nonumber\\
 &=\left(\sqrt{m^{2}+({\bf P}-{\bf q}/2)^{2}}-\sqrt{m^{2}+({\bf P}+{\bf q}/2)^{2}}\right)^{2}-{\bf q}^{2}\nonumber\\
&=-{\bf q}^{2}+  \left(\frac{{\bf P}\cdot {\bf q}}{m}\right)^{2}+{\cal O}( 1/m^{4})\,,
\label{eq:energy}
\end{align}
so
\begin{equation}
F(q^{2})=e^{-\frac{1}{6}{\bf q}^{2}\Delta^{2}}\left(1+\frac{\Delta^{2}({\bf P}\cdot{\bf q})^{2}}{6m^{2}}+{\cal O}(1/m^{4})\right)\,.
\label{eq:ffcorr}
\end{equation} 
Substituting eq.~(\ref{eq:ffcorr}) and (\ref{eq:energy}) into $\rho(r)$, we obtain
\begin{align}
\label{eq:rho-expanded}
\rho(r)&\cong\left(\frac{2R^{2}}{3\pi}\right)^{3/2}\!\!\!\int \!\!\frac{d^3P\, d^3q}{(2\pi)^3 } \left(1 + \left(\tfrac{1}{2m^{4}}+\tfrac{\Delta^{2}}{6m^{2}}\right) ({\bf P}\cdot {\bf q})^{2}\right)\nonumber\\
&\times \exp{ \left(-\tfrac{1}{6}{\bf q}^{2}\Delta^{2}-\tfrac{2}{3}{\bf P}^{2}R^{2} 
-\tfrac{1}{6}{\bf q}^{2}R^{2}+i{\bf q}\cdot{\bf r}\right)}\,.
\end{align}
All of these integrals can be performed analytically yielding
\begin{align}
\rho(r)&= \left(1+ \frac{27}{8}\left(\frac{1}{m^{4}}+\frac{\Delta^{2}}{3m^{2}}\right)\left(\frac{R^{2}+ \Delta^{2}-r^{2}}{R^{2}(R^{2}+ \Delta^{2})^{2}}\right)\right)\rho_{0}(r)\nonumber\\
&{\rm where}\nonumber\\
\rho_{0}(r) &=\left(\frac{3}{2\pi(R^{2}+ \Delta^{2})}\right)^{3/2}e^{-3r^{2}/2(R^{2}+ \Delta^{2})}\,.
\label{eq:rhofinal}
\end{align}  
The expansion in inverse powers of $m$ that made it possible to perform the integrals of \eq{eq:rho-expanded} limits the validity of this formula to parameter ranges where $m^{2}R^{2}$ and $m^{2}\Delta^{2}$ are large compared to one.  

The traditional identification of the charge density distribution with the Fourier transform of the form factor is obtained by first taking the mass $m$ to infinity and then taking the target wave packet radius to zero.  The result for the gaussian model is the ``naive'' charge distribution,
\begin{equation}
\rho_{\rm naive}(r)=\left(\frac{3}{2\pi \Delta^{2}}\right)^{3/2}e^{-3r^{2}/2\Delta^{2}}\, 
\label{eq:rhonaive}	
\end{equation}
with $r_{\rm RMS}=\Delta$, which is indeed the Fourier transform of the form factor, \eq{eq.ff-gauss}. This is the sequence of limits implicitly assumed by Sachs\cite{sachs}.  To get a feel for the dependence of \eq{eq:rhofinal} on the various parameters, it is useful to compute the mean-squared charge radius given by \eq{eq:rhofinal},
\begin{align}
\left< r^{2}\right> &=4\pi\int_{0}^{\infty}dr  r^{4}\rho(r)\nonumber\\
&=  \Delta^{2}\left(1-\frac{3}{4m^{2}R^{2}}\right) +R^{2}\left(1-\frac{9}{4m^{4}R^{4}}\right)\,.
\label{eq:msqr}
\end{align}
When the wave packet is large, the mean-squared charge radius is approximately $R^{2}$, the radius of the wave packet.  To obtain the traditional result, $<r^{2}_{\rm naive}>=\Delta^{2}$, it is necessary to choose $m^{2}R^{2}\gg 1$ in order to minimize the relativistic corrections in \eq{eq:msqr} and to choose $R^{2}\ll \Delta^{2}$ so that the wave packet dependent second term in \eq{eq:msqr} is negligible compared to the first.  Altogether the condition for a localization independent charge density is 
\begin{equation} 
\Delta^{2} \gg R^{2}\gg 1/m^{2}\,,
\label{eq:inequality}
\end{equation}
as quoted in \eq{eq:diffresult}.  If $\Delta^{2}$ is not much larger than $1/m^{2}$, then the charge density distribution in the system depends unavoidably on the wave packet used to localize it and \eq{eq:cmrel} must be evaluated numerically. 
 
\Eq{eq:inequality} is easily satisfied for atoms and large nuclei, so before going on to the most interesting case of the nucleon, we examine the  uncontroversial cases of atoms and heavy nuclei.

\section{Charge density distributions for atoms and nuclei}
\subsection{Atoms}
\begin{figure}
\begin{center}    
\includegraphics[width=8.6cm]{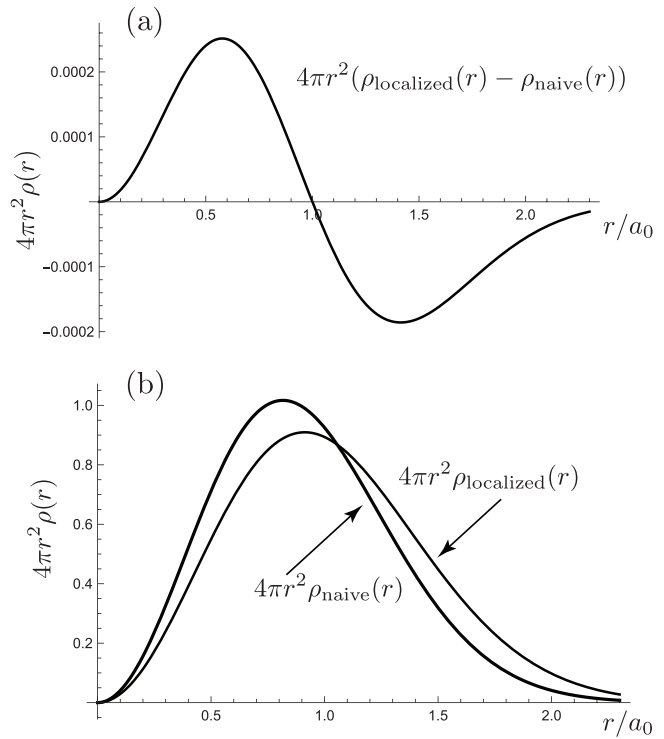} 
\caption{\small  Radial charge density distributions, $4\pi r^{2}\rho(r)$, in a   gaussian model for a hydrogen atom. (a) The difference between the charge density distribution for a hydrogen atom localized within 10~fm and the naive charge density distribution obtained by Fourier transform of its charge form factor. (b) The radial charge distribution of a gaussian hydrogen atom localized within $0.25\,a_{0}$ compared to  the naive charge distribution. }
\label{fig:hydrogen}
\end{center}
\end{figure}

The intrinsic size of atoms is roughly $a_{0}$, the Bohr radius, while their masses grow with $A$.  So atomic hydrogen is the worst case example among atoms.  
For atomic hydrogen, we take $\Delta = a_{0} \cong 5\times 10^{-11}$~m, and $m=m_{p}=1/(2.1\times 10^{-16}~\rm{m})$.  Since $\Delta \cong 2.5\times 10^{5}/m$ the inequalities of \eq{eq:diffresult} can be satisfied for a large range of $R$.  Explicit calculation shows that $\rho(r)$ is virtually indistinguishable from $\rho_{\rm naive}(r)$ for $R$ between 10~fm and $a_{0}/10$.  If we choose, for example, $R=10$~fm (corresponding to $mR\cong 50$, so the expansion of the previous section should be valid), Figure~\ref{fig:hydrogen}(a) shows the deviation of the radial charge density ($4\pi r^{2}\rho(r)$) of the localized atom from the naive result of \eq{eq:rhonaive} is less than  $\sim0.0002$ over whole range of $r$.  Thus we can conclude that to an accuracy of roughly 0.02\%, the Fourier transform of the form factor $F(-{\bf q}^{2})$ can be interpreted as a localization independent charge density distribution for the atom with a ``resolution'' of order $10^{-14}$~m, a distance that is of order 0.02\% of the intrinsic size of the system.  If, on the other hand, we take $R$ to be an appreciable fraction of the size of the atom, then the charge density distribution gets broader, reflecting the spread in the quantum wave packet.  Figure~\ref{fig:hydrogen}(b) illustrates this effect by comparing the naive charge density distribution with the localized distribution for $R=a_{0}/4$.
 
 We conclude that the naive charge density distribution obtained by Fourier transform of the form factor provides an excellent representation of the charge density distribution of a localized atom.  

\subsection{Nuclei}

\begin{figure}
\begin{center}  
\includegraphics[width=8.6cm]{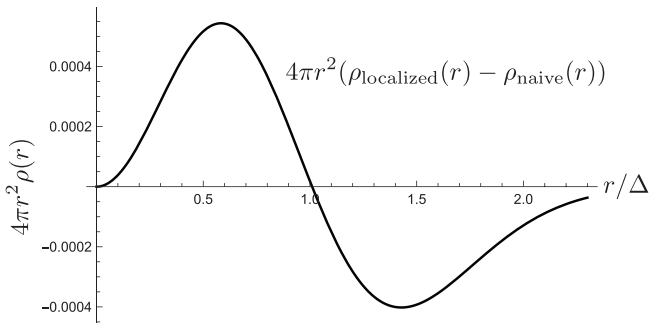} 
\caption{\small    The difference between the radial charge density distribution for a carbon nucleus localized within 0.2~fm and the naive radial charge density distribution obtained by Fourier transform of its charge form factor. }
\label{fig:carbon}
\end{center}
\end{figure}

For a typical nucleus we take $\Delta = 1.3 A^{1/3}$~fm and $m\cong Am_{p} 
\cong A/(.21~{\rm fm})$.  The constraint on $R$ becomes progressively easier to satisfy with increasing $A$.  Take carbon ($A=12$) for example (see below for the case of a very light nucleus such as deuterium) with $\Delta\cong 3$~fm.  Figure \ref{fig:carbon} shows the difference between the charge density distribution for carbon localized within  $R=0.2$~fm and the Fourier transform of the charge form factor.   The difference is less than 0.0005  over the entire range of $r$ indicating that the   association of the charge distribution with the Fourier transform of the form factor is valid to this accuracy with a resolution of order the nucleon's Compton wavelength, 0.2~fm.  Clearly the large mass of even a light nucleus such as carbon makes it possible to localize its center of mass to a region small compared to its intrinsic size.

To apply this analysis to hadrons and to the deuteron we must evaluate the integral of \eq{eq:cmrel} numerically.

\section{The charge density distribution in the nucleon and deuteron}

\begin{figure}
\begin{center}  
\includegraphics[width=8.6cm]{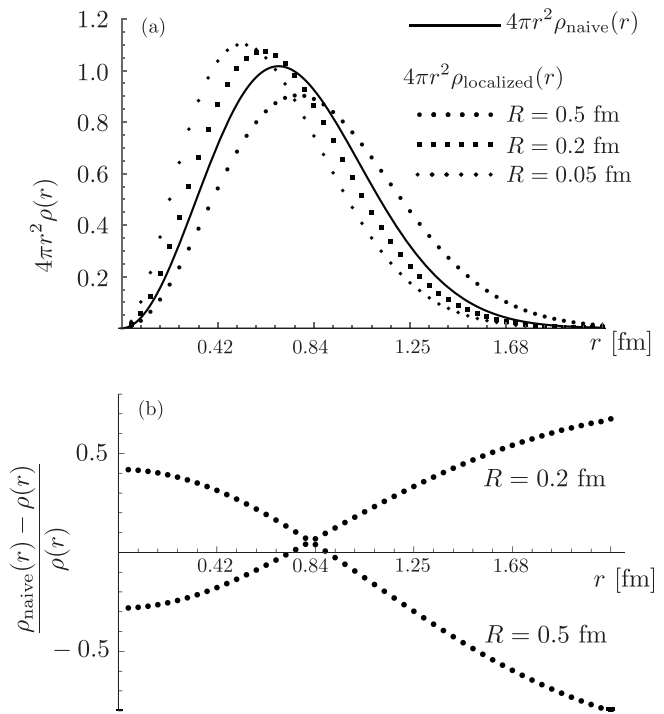} 
\caption{\small  (a) Radial charge density distributions, $4\pi r^{2}\rho(r)$, in a  gaussian model for the proton. The solid curve is the naive radial charge density distribution of \eq{eq:rhonaive}.  The dotted curves are the radial charge density distributions for a proton localized with $R=0.05,0.2$ and $0.5$ fm;  (b) The fractional difference between the naive and localized charge density distributions for $R=0.2$~fm and $R=0.5$~fm.}
\label{fig:proton}
\end{center}
\end{figure}

For the nucleon with $1/m\cong 0.2$~fm and $\Delta\cong 0.85$~fm the expansion $1/mR$ used in the previous section is not valid and \eq{eq:cmrel} must be evaluated numerically.  Figure~\ref{fig:proton} shows the results of trying of localize the model proton (with a gaussian form factor) at various distance scales.  In Figure~\ref{fig:proton}(a) we show the charge density for the proton localized with $R=0.5, 0.2$, and $0.05$~fm from \eq{eq:cmrel} compared with the naive charge density distribution of \eq{eq:rhonaive}.  The fact that for all values of $R$ the charge density distribution has the same qualitative shape as the naive charge density distribution should not be surprising:  $4\pi r^{2}\rho(r)$ is positive definite, normalized to one, grows like $r^{2}$ at small $r$ and falls like a Gaussian at large $r$.  Thus the overall shape of $4\pi r^{2}\rho(r)$ is highly constrained.  Within those constraints the $R$ dependence is significant. To display the dependence more clearly we plot the fractional deviation,
$$
\frac{\rho_{\rm naive}(r)-\rho(r)}{\rho(r)}\,,
$$ 
for $R=0.2$ and $0.5$~fm.  The fractional differences are of order one, and change dramatically between the two values of $R$, indicating that the concept of a \emph{localization independent charge density distribution for the proton is not well-defined}.

\begin{figure}
\begin{center}  
\includegraphics[width=8.6cm]{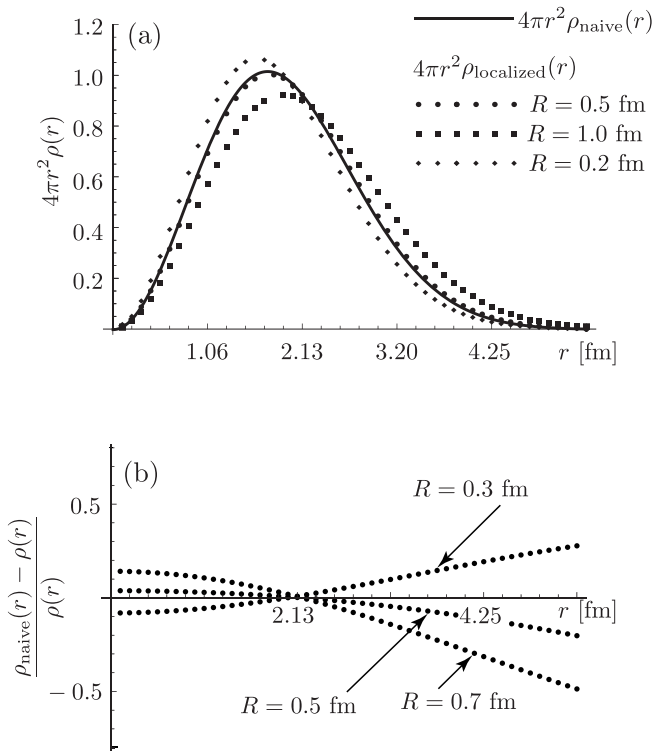} 
\caption{\small  (a) Radial charge density distributions, $4\pi r^{2}\rho(r)$, in a gaussian model for the deuteron. The solid curve is the naive radial charge density distribution of \eq{eq:rhonaive}.  The dotted curves are the radial charge density distributions for a deuteron localized with $R=0.2,0.5$ and $1.0$ fm;  (b) The fractional difference between the naive and localized charge density distributions for $R=0.3$~fm and $R=0.5$~fm and $R=0.7$~fm.}
\label{fig:deuteron}
\end{center}
\end{figure}

The deuteron presents an intermediate case.  Its intrinsic size is of order $\Delta \cong 2.15$~fm as determined by the slope of its charged form factor at $q^{2}=0$.  The deuteron Compton wavelength is $1/m\cong 0.1$~fm, so we expect that the naive charge density distribution should approximate the charge density distribution of a deuteron with values of between these two.  Explicit calculation shows that this is the case for values of $R$ near 0.5~fm, roughly the geometric mean of $m$ and $\Delta$.  This is illustrated in Figure~\ref{fig:deuteron}.  In particular,   Figure~\ref{fig:deuteron}(b) shows that the naive charge density agrees with the localized distribution to $\sim$ 20\% between $r=0$ and $\sim1.5\Delta$ for a range of values of $R$ around 0.5~fm.  The fractional difference increases at large $r$ where the charge density itself is small.  We conclude that the Fourier transform of the deuteron charge form factor gives a charge density distribution that provides a fair approximation to that of a deuteron localized at distances small compared to its intrinsic size.

The pion, an extreme example with an intrinsic size of order $\Delta\sim.5$~fm and a Compton wavelength of $1/m\cong 1.4$~fm, cannot be localized to distances of order its intrinsic size without generating relativistic effects that destroy the relationship between the Fourier transform of its form factor and its charge distribution.

\section{Discussion and conclusions}

A simple, Fourier transform relationship between form factors and spatial distributions of the expectation values of local operators was developed during the study of non-relativistic systems like atoms in the early days of quantum mechanics.  Although Burkardt\cite{Burkardt:2000za} and Miller\cite{Millerold} pointed out that this relationship fails in the case of the nucleon, the relationship seems to have entered the folklore of particle physics without careful consideration of whether it is accurate for systems whose size is of the same order as their Compton wavelength.
Miller, in particular, has emphasized that the connection fails for the nucleon\cite{Millerold} and has recently re-emphasized that it fails for the famous assertion that the mean squared charge radius of the proton is given by $-6G_{E}'(0)$\cite{Miller:2018ybm}.  Defining a quantity such as $\langle \hat \rho(r)\rangle_{N}$ requires one to localize the nucleon and doing so generates localization dependent contributions  
that invalidate the Fourier transform relation between form factors and local density distributions.
  I have explored this effect in the simple case for the charge density distribution of a spinless system with a gaussian form factor.  The problem is not special to the charge density operator nor to a spinless hadron nor to  the assumption of a gaussian form factor.  Instead this is a general problem that afflicts attempts to extract spatial distributions of local properties of any system that is not much larger than its Compton wavelength.  The problem is quite fundamental, since it originates in the interplay between the uncertainty principle and relativity.   

One can, of course, construct a function of ${\bf r}$ by Fourier transforming the form factor of a local operator, but in the case of the nucleon or other light hadrons, this is of uncertain value and should not be considered an accurate representation of the ``actual'' spatial distribution of the operator matrix element, which cannot be defined independent of the way in which the hadronic system was localized.

 \begin{acknowledgements}
 I am grateful Matthias Burkardt for conversations many years ago as well as more recently, to Dimitra Pefkou for recent conversations on this subject and for informing me of Jerry Miller's recent paper\cite{Miller:2018ybm}, and to Jerry Miller for very helpful comments on the manuscript.  I also appreciate comments and suggestions from Phiala Shanahan and Will Detmold.
\end{acknowledgements}

\

\end{document}